\documentclass[aps,preprint]{revtex4}
\usepackage{times}
\usepackage{graphicx}
\usepackage{natbib}
\usepackage{dcolumn}
\usepackage{psfig}
\makeatletter
\def\@cite#1#2{$(#1\if@tempswa , #2\fi)$}
\makeatother
\begin{document}
\begin{flushleft}
 {\Large Physical Sciences (Physics)} \\
\vspace {1.0in}
  {\LARGE\textbf{\textsf{Effects of lengthscales and attractions on
  the collapse of hydrophobic polymers in water }}}
  
  {\textbf{\textsf{{Manoj V. Athawale,$^*$  Gaurav Goel,$^\dag$ 
          Tuhin Ghosh,$^*$ Thomas~M.~Truskett,$^\dag$ and 
          Shekhar Garde$^*$ }}}}
  
  {\footnotesize{$^*$The Howard P. Isermann Department of Chemical \&
  Biological Engineering and Center for Biotechnology \&
  Interdisciplinary Studies, Rensselaer Polytechnic Institute, Troy,
  NY 12180;\\ $^\dag$Department of Chemical Engineering and Institute
  for Theoretical Chemistry, University of Texas,\\ Austin, TX
  78712.\\}}

{\underline {Corresponding Authors}} : Shekhar Garde;~e-mail
:gardes@rpi.edu. Thomas M. Truskett;~e-mail :truskett@che.utexas.edu.\\

\vspace {0.25in}

 {\underline {Number of pages}} : 18 (text), 5 (figures)

\vspace {0.25in}

 {\underline {Words in the abstract}} : 192

\vspace {0.25in}

 \underline {Total characters in the manuscript} : $\sim$ 46000




\end{flushleft}

\clearpage 

\centerline {\textbf {\textsf {ABSTRACT}} } 

\vspace {0.25in}
We present results from extensive molecular dynamics simulations of
collapse transitions of hydrophobic polymers in explicit water focused
on understanding effects of lengthscale of the hydrophobic surface and
of attractive interactions on folding. Hydrophobic polymers display
parabolic, protein-like, temperature-dependent free energy of
unfolding. Folded states of small attractive polymers are marginally
stable at 300 K, and can be unfolded by heating or cooling.
Increasing the lengthscale or decreasing the polymer-water attractions
stabilizes folded states significantly, the former dominated by the
hydration contribution.  That hydration contribution can be described
by the surface tension model, $\Delta G=\gamma (T)\Delta A$, where the
surface tension, $\gamma$, is lengthscale dependent and decreases
monotonically with temperature.  The resulting variation of the
hydration entropy with polymer lengthscale is consistent with
theoretical predictions of Huang and Chandler ({\it
Proc. Natl. Acad. Sci.}, {\bf 97}, 8324-8327, 2000) that explain the
blurring of entropy convergence observed in protein folding
thermodynamics. Analysis of water structure shows that the
polymer-water hydrophobic interface is soft and weakly dewetted, and
is characterized by enhanced interfacial density fluctuations.
Formation of this interface, which induces polymer folding, is
strongly opposed by enthalpy and favored by entropy, similar to the
vapor-liquid interface.

\newpage
\section{Introduction}
\vspace{-0.4cm} Hydrophobic interactions are one of the major
contributors to biological self-assembly in solution, including
protein folding and aggregation, micelle and membrane formation, and
biomolecular recognition
\cite{kauzmann:59:apc,tanford:73:book,dill:90:biochem,chandler:05:nature, 
pratt:02:arpc}. Recent work in this area has focused on the
lengthscale dependencies of hydrophobic hydration and interactions
\cite{lum:99:jpc,huang:00:pnas,chandler:02:nature,chandler:05:nature, 
rajamani:05:pnas}. In particular, a recent theory by Lum, Chandler,
and Weeks (LCW) \cite{lum:99:jpc} highlighted the different physical
mechanisms of solvation of small and large hydrophobic solutes in
water.  Small solutes are accomodated in water through molecular-scale
density fluctuations \cite{pratt:77,hummer:96:pnas}, whereas solvation
of larger solutes requires formation of an interface similar to that
between a liquid and a vapor \cite{chandler:05:nature,
lum:99:jpc,stillinger:73:jsc}. This change in physics is also
reflected in thermodynamic (entropy vs enthalpy dominated hydration)
\cite{rajamani:05:pnas} and structural (wetting vs dewetting of the
solute surface) \cite{chandler:05:nature,
hummer:98:prl,stillinger:73:jsc} aspects of hydration.  Similarly,
interactions between larger hydrophobic solutes in water
\cite{wallqvist:95:jpc1, wallqvist:95:jpc2,berne:03:pnas,
choudhury:05:jacs,smith:05:pre} are characteristically distinct from
those between their molecular counterparts
\cite{smith:93:jcp,ghosh:02:jcp}.

The differences between the hydration and interactions of small and
large solutes characterize many-body effects in hydrophobic phenomena.
Effects of similar origin are also at work in association of small
hydrophobic solutes into a larger aggregate \cite{levitt:97:proteins,
levitt:01:pnas}, and are quantified by the $n$-particle potential of
mean force (PMF) \cite{ghosh:03:jpcb, shimizu:02:protein,
scheraga:05:jpcb1,baker:97:protein}. 
For $n>3$ however, the dimensionality of the system makes calculations
of $n$-particle PMFs computationally prohibitive.

To this end, ten Wolde and Chandler \cite{tenwolde:02:pnas} performed
studies of a hydrophobic polymer in water, a many-body model system of
potential relevance to protein folding.  Their calculations, using a
coarse-grained model of water, showed that a sufficiently long
hydrophobic polymer readily collapses in water driven by a significant
free energy difference ($\sim$30$k_BT$ for their model polymer). Their
simulations also predicted that the essentially wet extended states of
the polymer go through dewetted transition states at a smaller radius
of gyration before collapsing into an ensemble of compact globular
states.
Although molecular simulation studies of conformational equilibria of
solvophobic oligomers in explicit water and in other solvents have
been performed in the past \cite{wallqvist:96:bj,ashbaugh:99:bj,
johnston:96:jcp,ghosh:05:jpcb, paschek:05:pccp,siepmann:jpcb:06}, a
systematic study of how attractive interactions and lengthscales
affect the thermodynamics of collapse or folding-unfolding transitions
has not been reported.

Motivated by the work of ten Wolde and Chandler
\cite{tenwolde:02:pnas}, here we present results from extensive
molecular dynamics simulations that address the dependence of folding
of hydrophobic polymers in water on lengthscales and attractive
interactions.  We show that compact folded states of smaller polymers
with attractive polymer-water interactions are only marginally stable
and display thermodynamic characteristics of both warm and cold
denaturation. Increasing the polymer lengthscale makes the hydration
contribution to folding more favorable, and stabilizes folded states.
Reducing the polymer-water attractions also stabilizes folded
states. We explore the suitability of area and volume based models to
describe the hydration contribution to folding. Entropy obtained from
temperature dependence of the hydration contribution to folding varies
with the polymer lengthscale, explaining the origin of blurring of the
`entropy convergence' in protein folding using arguments proposed by
Huang and Chandler \cite{huang:00:pnas}. Structural signatures, such
as water density and its fluctuations in the vicinity of the polymer,
indicate that the polymer-water hydrophobic interface is soft and
weakly dewetted and allows enhanced density fluctuations, collectively
inducing a collapse into compact folded conformations.

\noindent {\textbf {\textsf {RESULTS AND DISCUSSION}} } 

To observe collapse transitions potentially important to understanding
protein folding thermodynamics, the model polymer needs to be
sufficiently long, such that it can adopt compact globular as well as
extended states.  Preliminary studies indicate that 6-, 8-, or 12-mer
chains are too short to form compact states. In contrast, somewhat
longer, 25-mer hydrophobic oligomers, can fold into tightly wound
helical states.  We therefore focus on the collapse transitions of
such 25-mer hydrophobic chains in water.

To study the lengthscale dependence, we studied two versions of the
25-mer, a smaller lengthscale polymer (denoted {\bf C25}) comprising
methane-sized monomers ($m$) (with $\sigma_{mm}$=0.373 nm and $m-m$
bond length of 0.153 nm, and $m-m-m$ bond angle of 111.0 deg), and a
larger lengthscale polymer (denoted {\bf CG25}) comprising larger
(ethane-sized) monomers ($M$) (with $\sigma_{MM}=$0.44 nm, and $M-M$
bond length of 0.25 nm). The $M-M-M$ bond angle potential was not
employed in simulations of the larger lengthscale CG25 polymer, and
dihedral potentials were not employed for either polymers.  We note
that we are not interested in detailed studies of specific alkanes in
water, but in conformational transitions of flexible hydrophobic
chains driven by hydrophobic interactions, and their dependence on
lengthscales, attractions, and temperature. The inclusion of
additional intrapolymer interactions ({\it e.g.}, a dihedral
potential) changes the conformational preferences but will not affect
the qualitative hydration contributions or their temperature
dependence that are of primary interest here.  Further, the larger
lengthscale polymer, CG25, can be thought of approximately as a
coarse-grained equivalent of a 50-mer of methane-sized monomers. The
$M-M-M$ angle potential was therefore turned off to capture the higher
conformational flexibility of that polymer.

To quantify effects of polymer-water interactions on collapse
transitions, we employed three different interactions for C25 and CG25
polymers: (i) Full LJ, where monomers interact with other nonbonded
monomers and water {\it via} Lennard Jones (LJ) interactions
[$\sigma_{mw}=(\sigma_{mm}+\sigma_{ww})/2$ and
$\epsilon_{mw}=\sqrt{\epsilon_{mm}\epsilon_{ww}}$].  We used
$\epsilon_{mm}=$ 0.5856 kJ/mol and $\epsilon_{MM}$=0.85 kJ/mol,
respectively. The strength of attractions employed here is similar to
that of typical alkane-water attractive interactions
\cite{jorgensen:opls}.  (ii) Half LJ, where monomer-monomer
$\epsilon_{mm}$ was reduced to half its value in (i), and
monomer-water $\epsilon_{mw}$ correspondingly by a factor of
$\sqrt{2}$, and, (iii) WCA, where Weeks-Chandler-Andersen \cite{wca}
repulsive potential was used to describe monomer-monomer and
monomer-water interactions.  We denote the full LJ and WCA versions of
the two polymers by {\bf C25-LJ}, {\bf C25-WCA}, {\bf CG25-LJ}, and
{\bf CG25-WCA}, respectively.

Figure 1 shows the potential of mean force (PMF), $W(R_g)$, for
conformational sampling of polymers along the radius of gyration
reaction coordinate obtained from umbrella sampling simulations (see
Methods).  Several interesting features are apparent.  For the C25-WCA
polymer, tightly wound helical configurations (with $R_g\sim 0.42 $nm)
are stabilized by $\sim$20 kJ/mol relative to extended configurations
in the unfolded basin. Although the PMF decreases monotonically as the
polymer folds, the slope is not uniform -- PMF is relatively flat
beyond $R_g$ of 0.65 nm and rapidly decreases below that value.

For the CG25-WCA polymer comprising larger monomer units, the driving
force for folding is significantly higher.  The folded states are
stabilized by almost 100 kJ/mol relative to extended states. Indeed,
in absence of the umbrella potential, the extended states of this
polymer collapse rapidly into compact states over a timescale of
$\sim$100 ps. Once folded, no significant partial or complete
unfolding events are observed indicating the strength of hydrophobic
driving force for the polymer of this size.

To what extent do attractive interactions change the folding-unfolding
of hydrophobic polymers in water?  Figure 1 also shows PMF profiles
for C25-LJ and CG25-LJ polymers interacting with full LJ potential
having magnitude similar to typical alkane-water attractions. Here,
the attractive interactions directly affect two opposing enthalpic
terms -- the monomer-monomer attractions favor folding of the polymer
into compact states in absence of other effects (Fig. 2c), whereas
monomer-water attractions favor unfolding (Fig. 2b) to increase the
favorable energetic interactions with water.

For the smaller, C25-LJ polymer, balance of these interactions leads
to two distinct minima corresponding to folded and unfolded ensembles
separated by a barrier.  Overall, the polymer-water attractive
interactions dominate over other contributions, leading to a slight
stabilization of the unfolded states compared to the folded ones.  In
contrast, for the larger CG25-LJ polymer, despite the significant
polymer-water attraction contribution favoring the unfolded states
(Fig. 2b), the qualitative nature of the PMF is similar to that of
CG25-WCA polymer -- folded states are favored significantly over the
unfolded states.

The total PMF, $W(R_g)$, governs the overall conformational behavior
of polymers in water, and can be separated into three contributions
\cite{tenwolde:02:jpcm}:
\begin{equation}
\label{eqn:pmf}
W(R_g) = W_{vac}(R_g) + \left< U_{PW}(R_g) \right> + W_{hyd}(R_g),
\end{equation}
where $W_{vac}$ is the PMF profile in vacuum, $\left< U_{PW} \right>$
is the ensemble averaged polymer-water interaction energy (for given
$R_g$), and we define $W_{hyd}$ as the hydration contribution.  We
note that $[W(R_g)-W_{vac}(R_g)]$ is traditionally referred to as the
`solvent contribution'.  However, its separation in Eq. 1 into $\left<
U_{PW}(R_g) \right>$ and $W_{hyd}(R_g)$ is similar to that used in
perturbation theory approaches applied previously to solvation
phenomena \cite{pratt:77,garde:99:bpc}.  The rationale for writing
Eq. \ref{eqn:pmf} is that each individual term may be separately
predicted using theoretical arguments or simplified models
\cite{tenwolde:02:jpcm}, as we illustrate below for the 
$W_{hyd}(R_g)$ term.  For nonpolar solutes in water, the solute-water
attractive energy is generally smaller in magnitude than the hydration
contribution, whereas for polar or ionic solutes, it will dominate the
solvation free energy.  In either case, a similar separation may
provide a useful predictive framework.


Figure 2 shows $W_{hyd}(R_g)$, $\left< U_{PW}(R_g) \right>$, and
$W_{vac}(R_g)$ contributions to $W(R_g)$ for LJ and WCA versions of
the C25 and CG25 polymers.  For both the polymers, the hydration
contribution is large, approximately 40 kJ/mol and 140 kJ/mol,
respectively, and monotonically favors folding of polymers into
compact states.  Despite the significant attractive interactions for
LJ polymers, $W_{hyd}$ makes the dominating contribution to the
overall PMF.  More importantly, the $W_{hyd}(R_g)$ contribution
appears to be similar for all C25 or CG25 polymers (full LJ, half LJ,
or WCA), relatively insensitive to the strength of attraction.  This
supports the separation of the PMF in Eq. 1, and indicates that the
$W_{hyd}$ term captures the essential physics of hydrophobic hydration
that may be described either using cavity formation (for small
lengthscales) or interface formation process (for larger
lengthscales).


Once the overall $W(R_g)$ profile is available from simulations, the
total free energy of unfolding a polymer in water, $\Delta G_u$, can
be obtained by integration of that PMF \cite{hill:56:book}. Figure 3
shows the temperature dependence of $\Delta G_u$, calculated in this
manner.  Effects of both lengthscale and attractive interactions on
the free energy of unfolding are apparent.  For the smaller and
attractive C25-LJ polymer, the $\Delta G_u$ profile is parabolic with
compact states being only marginally stable (by $\sim$1 kJ/mol) for
$350<T< 450$ K.  Thus, this polymer shows signatures of both warm and
cold denaturation processes. Increasing the lengthscale increases the
hydrophobic driving force for collapse. Indeed, folded states of
CG25-LJ polymer are stabilized by over 10 kJ/mol at the temperature of
maximum stability.  Reducing the polymer-water attractions or removing
them completely also increases the driving force for folding. For
example, at 300 K, the $\Delta G_u$ values for C25-WCA and CG25-WCA
polymers are about 10 and 30 kJ/mol, respectively.

The hydration contribution, $\Delta G_u^{hyd}$, to the total free
energy of unfolding estimated using $W_{hyd}(R_g)$ is shown in Figure
4a.  Several features of these curves are noteworthy. The $\Delta
G_u^{hyd}$ is positive at all temperatures, is larger in magnitude for
the CG25 polymers, and decreases monotonically with increasing
temperature for all polymers.  Because the polymer-water attractions
are not included in this term, the $\Delta G_u^{hyd}(T)$ profile is
similar and favors folding of both WCA and LJ polymers.  Figures 3 and
4 together show that for the purely repulsive WCA polymers, $\Delta
G_u^{hyd}$ determines the temperature dependence of the overall
$\Delta G_u$, whereas for the attractive LJ versions of the polymer,
the additional polymer-water energy contribution (which favors
unfolding) makes the $\Delta G_u(T)$ profile parabolic.

The hydration contribution quantifies the difference in hydration free
energy of folded and unfolded state ensembles, $\Delta
G_u^{hyd}(T)=G_u^{hyd}(T)-G_f^{hyd}(T)$.  
Thus, to predict $\Delta G_u^{hyd}(T)$, we can (I) explore the
applicability of simpler models that predict free energies of
hydration of representative folded and unfolded states using
lengthscale-dependent physics of hydration; or (II) attempt to model
the entire conformation-dependent potential of mean force
$W_{hyd}(R_g)$, and obtain $\Delta G_u^{hyd}$ by appropriate
integration.  Below we explore path (I) first.

The folded globular states of C25 and CG25 have equivalent radius
[$(3V/4\pi)^{1/3}$ or $(A/4\pi)^{1/2}$] of approximately 0.65 and 0.95
nm, respectively; where $V$ is the solvent-excluded volume (SEV), and
$A$ is the solvent-accessible surface area (SASA) of the polymer.
These radii values are in the `crossover' region
\cite{rajamani:05:pnas, chandler:05:nature}, and we expect the hydration 
free energy of the folded state to be adequately described by the
surface tension model, $G_f^{hyd}=\gamma A$ \cite{tenwolde:02:jpcm}.
In contrast, the lengthscale of unfolded polymers is more difficult to
quantify.  If we assume the relevant lengthscale to be small, equal to
that of a monomer, then the free energy of hydration of unfolded
polymers will be roughly proportional to the solvent excluded volume
\cite{chandler:05:nature,huang:00:pnas}, approximated as $n
\times G_{mon}^{hyd}(T)$ (where $n$ is the number of monomers, 
and the subscript $mon$ indicates the value for a monomer), or more
quantitatively as, $V\times c(T)$, where the volume coefficient
$c(T)=G_{mon}^{hyd}/V_{mon}$ (or alternatively, $=\partial
G_{mon}^{hyd}/\partial V_{mon}$).

Figure 4b shows the hydration contribution to the free energy of
unfolding predicted using the `area-volume' model, $\Delta
G_u^{hyd}(T) = V \times c(T) - \gamma (T) A$.  We obtained $V$ and $A$
values using the molecular surface package of Connolly
\cite{connolly:msp}.   Values of $c(T)$ ($\sim$150-170 kJ/mol/nm$^3$) 
obtained independently from test particle insertions of C25 and CG25
monomers into pure water systems, are in excellent agreement with that
used by ten Wolde recently \cite{tenwolde:02:jpcm}.  The surface
tension, $\gamma (T)$, for folded states can be approximated by that
of an equivalent sphere.  These data are difficult to obtain by test
particle insertions \cite{rajamani:05:pnas}, but can be estimated
using the data in Figure 2 in Huang and Chandler \cite{huang:00:pnas}
(as $\Delta \mu /A$), or by using Tolman corrected macroscopic surface
tension, $\gamma (T) = \gamma_\infty (T) (1-2\delta (T)/R)$ (where the
Tolman length, $\delta (T)$ is taken from \cite{pratt:06:rmp}, and
$\gamma_\infty (T)$ is the surface tension of water
\cite{spce:tension}), or by using $\gamma (T) = [\partial W_{hyd} /
\partial A]_T$.  Figure 4b shows the prediction using the area-volume
model, where $\gamma (T)= [\partial W_{hyd} / \partial A]_T$ was used.
The prediction not only differs numerically from simulation values of
$\Delta G_u^{hyd}$, but shows opposite temperature dependence.  We
find that the other two methods of estimating $\gamma (T)$ give
similar results.


Why does the physically motivated area-volume model not provide a good
description of $\Delta G_u^{hyd}$ obtained from simulations?  This
model attempts to predict the temperature-dependent excess chemical
potentials of hydration (or the vacuum-to-water transfer free
energies) of folded and unfolded states independently -- a truly
challenging task. We note that only the difference $\Delta
G_u^{hyd}(T)$, and not these independent values are currently
available from simulations. Therefore, we can only make qualitative
arguments here regarding the applicability of the area-volume model to
the present systems.
For example, the unfolded states of hydrophobic polymers comprise
monomers joined together by covalent bonds, as distinct from fully
separated and well dispersed monomers in solution. Also, there is
conformational diversity in the unfolded states that includes fully as
well as partially extended conformations with different local
curvatures.  A single small monomer-like lengthscale therefore does
not appear to describe the hydration free energy of unfolded states,
and especially, its temperature dependence.  In contrast, many other
processes do involve assembly of small lengthscale moieties into a
larger lengthscale aggregate ({\it e.g.}, in micelle formation
\cite{maibaum:04:jpcb}) where the area-volume model may provide a 
good description of hydrophobic contributions to the thermodynamics of
assembly.

We also explored the suitability of alternate models (i) $\Delta
G_u^{hyd}(T) = c(T) \Delta V_u$ and (ii) $\Delta G_u^{hyd}(T) =\gamma
(T) \Delta A_u$, where $\Delta V_u$ and $\Delta A_u$ are the volume
and areas of unfolding, respectively, determined from our simulations.
Here, volume or area-based description alone is assumed to apply to
both folded and unfolded state ensembles.
We find that the volume-based calculation [model (i)] that uses $c(T)$
obtained from thermodynamics of small solute transfer significantly
underpredicts the $\Delta G_u^{hyd}$ at room temperature, as well as
its temperature dependence (not shown).  In contrast, the area-based
description [model (ii)] provides a quantitative prediction of $\Delta
G_u^{hyd}(T)$ as shown in Figure 4a.  $\gamma (T)$ constitutes a
critical input to this model, and was obtained using $\gamma (T) =
[\partial W_{hyd}/\partial A]_T$ for both polymers,
The numerical value of $\gamma$ is larger for CG25 polymer compared to
that for the smaller lengthscale C25 polymer.  Temperature dependence,
$\gamma (T)$, for both polymers is qualitatively similar to that of
vapor liquid interfacial tension measured in experiments or in
simulations of the SPC/E water \cite{spce:tension}.

In contrast to the area-volume model, the models (i) or (ii) used
above based on volume or area alone do not attempt to predict the
excess chemical potentials of folded or unfolded states independently,
but implicitly assume that the PMF $W_{hyd}$ varies linearly with
volume or area over the entire $R_g$ range.
How good is the assumption of linearity?  Figures 4c and d show
$W_{hyd}$ plotted as a function of $V$ and $A$, respectively.  All
values are referenced to their respective values at $R_g^{cut}$.
For the CG25-WCA polymer, variation of $W_{hyd}$ is not linear but
roughly sigmoidal with respect to $V$.  The slope, $c=\partial
W_{hyd}/\partial V$ is more than twice that of $c_{mon}$ obtained from
small solute studies ($\sim 171$ kJ/mol/nm$^3$) and approaches
$c_{mon}$ only at the ends. For the CG25-LJ polymer, the behavior is
similar, if somewhat less non-linear.  As the effective lengthscale of
the polymer is reduced, the slope decreases, and for C25-LJ polymer
approaches the $c_{mon}(298 K)$ value.  Inspection of these curves
over a range of temperatures shows a similar behavior.  For C25
polymers, even though $W_{hyd}$ is approximately linear with $V$ at
all temperatures, the slope $c(T)$ is different from $c_{mon}(T)$,
explaining the failure of volume-based $\Delta
G_u^{hyd}(T)=c_{mon}(T)\Delta V_u$ model (i) above.

Figure 4d shows that $W_{hyd}$ varies roughly linearly with area. The
slope, $\gamma = \partial W_{hyd}/\partial A$, is higher for CG25
polymers and decreases (from 50 dynes/cm to 25 dynes/cm) as the
effective lengthscale is reduced from CG25-WCA to CG25-LJ to C25-WCA
to C25-LJ. The temperature dependence of $\gamma$ for C25 and CG25
polymers is apparent from Figure 4a (as $\Delta A_u$ values are
constant). $\gamma$ decreases monotonically over the temperature range
of interest qualitatively similar to the vapor-liquid surface tension
of water.  In absence of detailed simulation data, how one may predict
{\it a priori} the exact numerical value of $\gamma (T)$ for a polymer
of a given lengthscale is not entirely clear.  Prediction using the
Tolman equation (typically used for spherical solutes) provides a
qualitatively correct picture, but is not quantitative (not shown).
Application of LCW theory \cite{lum:99:jpc,tenwolde:02:pnas} to
explore lengthscale and temperature dependence of $\gamma$ for
non-spherical solutes may be one computationally efficient
alternative.  Nevertheless, the approximate linearity with area and
the corresponding temperature dependence of the slope explains the
sucess of $\Delta G_u^{hyd}=\gamma (T)\Delta A_u$ model.


The free energies $\Delta G_u(T)$ and $\Delta G_u^{hyd}(T)$ obtained
here for hydrophobic polymers have important connections with the
thermodynamics of protein folding.  The $\Delta G_u^{hyd}(T)$ is
positive, favors folding, and decreases in magnitude with increasing
temperature, a trend similar to that of $\Delta G_u(T)$ for WCA
polymers.  The polymer-water attractions in contrast favor unfolding,
and more so at lower temperatures. Inclusion of attractions therefore
reduces the stability of folded states and makes the $\Delta G_u(T)$
curve roughly parabolic in T, similar to that for proteins.  To
summarize, increasing the lengthscale increases the magnitude of
$\Delta G_u(T)$ and decreases the temperature, $T^*$ of maximum
$\Delta G_u$, whereas addition of attractive interactions has the
opposite effect.

%

These observations have two important consequences. First, the
increased stability of folded polymers with reduction of polymer-water
attractive interactions as well as the shift of $T^*$ to lower values
with increasing lengthscale will make it difficult to observe cold
denaturation over temperature range accessible to experiments, without
assistance from other denaturing influences
\cite{privalov:92:bc,tanaka:02:bba}. Second, because the magnitude 
and temperature dependence of $\Delta G_u^{hyd}$ (and therefore, of
$\Delta G_u$) is sensitive to lengthscales, the corresponding
temperature-dependent entropy contributions will also vary with
polymer (or protein) size.  Sufficient variation in protein
lengthscales will lead to blurring of the ``entropy convergence''
observed in thermal unfolding of globular proteins
\cite{huang:00:pnas,garde:96:prl,baldwin:92:pnas,privalov:79:apc, 
ashbaugh:jcp:02}.

Figure 4e shows that for a methane-sized solute, the hydration
entropy, $\Delta S^{hyd}_u=-(\partial \Delta G_u^{hyd}/\partial T)$,
is large and negative at room temperature, characteristic of
hydrophobic hydration at small lengthscales
\cite{lazaridis:92:jpc}.  For C25 and CG25 polymers the $\Delta 
S_u^{hyd}$ is positive over the entire range of T, with its magnitude
being larger for the CG25 polymer. These observations are
qualitatively consistent with predictions by Huang and Chandler
\cite{huang:00:pnas}, who argued that such lengthscale dependence
of entropy explains the blurring of the `entropy convergence' observed
in unfolding of globular proteins. Also, the hydration contribution to
the heat capacity of unfolding, $\Delta C_{p-u}^{hyd}=T(\partial
\Delta S_u^{hyd}/\partial T)$, is positive, similar to that for
proteins.

Is the folding or collapse of hydrophobic polymers observed here
induced by the formation of a vapor-liquid like interface?  In their
study, ten Wolde and Chandler \cite{tenwolde:02:pnas} found that as a
hydrophobic polymer folds to more compact states, it goes through dry
or dewetted transition states with low vicinal water densities.  In
Figure 5, we quantify the vicinal water density by calculating number,
$N_w$, of water molecules in the first hydration shell divided by the
polymer SASA. These local density values are plotted as a function of
the polymer SASA and are normalized to 1 in the extended states of
polymers.

The normalization factor, equal to the vicinal density in the extended
state, is 14.1 for C25-LJ, 13.2 for C25-WCA, 13.3 for CG25-LJ, and
12.8 for CG25-WCA polymers (all in units of molecules/nm$^2$).  For
isolated C25 and CG25 monomers, the local surface densities obtained
from independent simulations are essentially identical for LJ and WCA
versions, equal to approximately 16 and 15 molecules/nm$^2$,
respectively. Thus, by this criterion, the extended states of polymers
themselves are already fairly dewetted compared to isolated monomers,
highlighting the differences between fully separated monomers and
unfolded states of polymers.

Folding of these polymers leads to further dewetting.  The extent of
that dewetting is rather small, $\sim$3 \% for the C25-LJ polymer.
When the attractive interactions are removed ({\it e.g.}, for the
C25-WCA case), the local density decreases, goes through a minimum
($\sim$10\% reduction) before reaching the folded states, for which
the reduction is about 6\%. This depletion is consistent with a small
increase in the lengthscale upon folding of the polymer. These effects
are enhanced for the larger CG25 polymers, which show density
depletion even for attractive polymer-water interactions (Figure 5b).

The fluctuations of water density in the hydration shell are also
noteworthy.  Figures 5c and 5d show normalized water number density
fluctuations in the first hydration shell, $\sigma^2_{N_w}\times {\rm
SASA}/\left< N_w \right> ^2$, which quantify approximately the
compressibility of the first hydration shell. (We note that
$\sigma^2_{N_w}$ also shows similar features). It is clear that for
the small attractive C25-LJ polymer, the hydration shell
compressibility changes little as it folds. In contrast, the larger
and repulsive CG25-WCA polymer shows increased absolute fluctuations
as well as a peak in the magnitude of fluctuations as the polymer
folds to compact states. Interestingly, although the nature of
$W(R_g)$ and the dewetting is qualitatively similar for CG25-WCA and
CG25-LJ polymers, an explicit peak in fluctuations is not observed
during folding of the CG25-LJ polymer.  This is expected as the
polymer-water attractions will dampen the longitudinal density
fluctuations.  In addition, solute-water attractions increase the
crossover lengthscale, or alternately, reduce the effective
lengthscale of the solute. In this sense, CG25-LJ lengthscale is
smaller than that of CG25-WCA polymer.


Collectively, for the polymers of the nanometer lengthscale studied
here, we find that the unfolded states are dewetted relative to
separated monomers, and the folding process is accompanied by further
dewetting of the polymer surface.  More importantly, the increased
density fluctuations for repulsive polymers indicate that the
polymer-water interface is soft, qualitatively similar to
water-hydrophobic liquid-liquid \cite{patel:03:jcp} or liquid-vapor
interfaces.  In addition, the positive hydration entropy indicates
that the unfavorable free energy is dominated by the enthalpic
contribution similar to that for interface formation
\cite{chandler:05:nature, rajamani:05:pnas}, supporting the above
assertion.  Extension of the above results would predict increased
dewetting, higher density fluctuations, and larger driving forces for
the folding of longer polymers.

\noindent {\textbf {\textsf {CONCLUSIONS}}}

We have presented results from extensive molecular dynamics
simulations of folding-unfolding of hydrophobic polymers in water with
focus on understanding effects of polymer lengthscale and
polymer-water attractive interactions on thermodynamic and structural
aspects of polymer folding.  The folding of model polymers provides a
route to characterizing the many-body hydrophobic effects in the
context of realistic biological self-assembly process, such as protein
folding.  We studied conformational transitions of 25-mers comprising
monomers of methane-like and larger particles.  To study the effects
of polymer-water attractions, we employed attractive LJ as well as
purely repulsive WCA interactions between monomers and water.  We
found that the smaller attractive polymer displays a parabolic
protein-like unfolding free energy, $\Delta G_u(T)$, profile as a
function of temperature. The folded states of this polymer are only
marginally stable at room temperature and can be unfolded by heating
or cooling the solution.  Increasing the lengthscale or reducing the
polymer-water attractions not only increases the stability of folded
states, but also shifts the temperature of maximum stability to lower
values.  Also, the dependence of hydration entropy on polymer
lengthscale is consistent with predictions of Huang and Chandler
\cite{huang:00:pnas} on lengthscale dependence of hydrophobic
hydration that provide a basis for understanding of the blurring of
entropy convergence observed in protein unfolding
thermodynamics.

The hydration contribution to folding of hydrophobic 25-mers studied
here is large and strongly favors compact folded states.  We explored
the applicability of physically motivated models based on area and/or
volume descriptions of hydration of polymer conformations to predict
the hydration contribution, $\Delta G_u^{hyd}$. Our analysis indicates
that predictions using $\Delta G_u^{hyd}=\gamma (T) \Delta A_u$ are in
excellent agreement with simulation data.  The surface tension
$\gamma$ is larger for CG25 polymers and decreases monotonically with
temperature for all polymers.  Structural details indicate that the
polymer-water interface, especially in the vicinity of the folded
state ensemble, is soft, weakly dewetted, and is characterized by
enhanced density fluctuations.  Correspondingly, $\Delta G_u^{hyd}$ is
enthalpy-dominated consistent with interface-driven folding process.

The polymer models presented here or the ones used by ten Wolde and
Chandler, present promising model systems of potential importance in
understanding the role of hydration in thermodynamics of protein
folding and aggregation \cite{berne:05:nature}.  The ability to change
intrapolymer and polymer-water interactions, and polymer size allows
for systematic studies of the free energy landscape in the extended
parameter space.  Further, investigation of the folding-unfolding of
these polymers over a range of solution conditions, such as high
pressures, in presence of cosolutes and solvents
\cite{ghosh:05:jpcb,paschek:05:pccp}, will provide important insights
into protein stability over the broader thermodynamic space.

\noindent {\textbf {\textsf {METHODS}} }

{\bf Intra polymer interactions:} The intra polymer potentials for
bond lengths in C25 and CG25 polymers and bond angles in C25 polymers
were harmonic in nature with following parameters: For $m-m$ bond
length in C25 polymers, $V_b = 0.5k_b(r - r_0)^2$, where $k_b$ = {\rm
334720.0} kJ/mol/nm$^2$ and $r_0$ = {\rm 0.153} nm, and for $m-m-m$
bond angle, $V_{\theta} = 0.5k_{\theta}(\theta - \theta_0)^2$, where
$k_{\theta}$ = {\rm 462.0} kJ/mol/deg$^2$ and $\theta_0$ = 111.0
deg. Similarly, for $M-M$ bond length in CG25 polymers, we used $k_b$
= {\rm 60702.0} kJ/mol/nm$^2$ and $r_0$ = {\rm 0.25}.

{\bf Free energy calculations:} We used the umbrella sampling technique
\cite{frenkel:book} to characterize the conformational free energy of
hydrophobic polymers in water as a function of polymer radius of
gyration, $R_g$. A harmonic potential, $\frac{1}{2}k_u(R_g-R_{g0})^2$,
was applied to efficiently sample polymer conformations in a window
near $R_{g0}$. For C25 polymers, we used 13 equally spaced windows
spanning a range of $R_g$ between the most compact ($\sim$ 0.37 nm)
and the most extended ($\sim$ 0.92 nm) states, whereas for CG25
polymers, we used 20 equally spaced windows with $R_g$ ranging from
$\sim$ 0.45 nm to $\sim$ 1.50 nm. Simulations of C25 polymers included
1700 water molecules (SPC/E model \cite{berendsen:87:jpc}). For CG25
polymers, depending on $R_{g0}$ value, the number of water molecules
ranged from 4000-12000. Additional simulations in which 12000 water
molecules were used uniformly in all windows confirmed that the
results presented here are insensitive to system size.

The values of $k_u$ ranged between 4000 and 12000 kJ/mol/nm$^2$, based
on the local gradient of the free energy, $dW(R_g)/dR_g$, estimated
from preliminary simulations.  Data from simulation trajectories -- 26
ns for the C25 polymer and 40 ns for the CG25 polymer (2 ns/window) --
were combined using the WHAM formalism
\cite{ferrenberg:89,kumar:92,roux:95} to generate a potential of mean
force profile along the $R_g$ coordinate for each polymer. Similar
calculations were performed in vacuum to obtain the PMF in the absence
of solvent.

{\it Simulation details:} MD simulations were performed using GROMACS
\cite{berendsen:95:cpc,lindahl:01:jmm} , with suitable modifications
to implement umbrella sampling as well as the WCA interaction scheme.
Periodic boundary conditions were applied and the particle mesh Ewald
method \cite{pme:93} was used to calculate the electrostatic
interactions with a grid spacing of 0.1~nm. NPT simulations were
performed at 1 atm and 300 K maintained using Berendsen algorithm
\cite{berendsen:84}.  The SETTLE algorithm \cite{settle} was used to
constrain OH and HH distances in water with a geometric tolerance of
0.0001~\AA.  Simulations were also carried out over a range of
temperatures to calculate temperature dependence of the free energy of
folding along the saturation curve of water \cite{garde:96:prl}.

\vspace{1.0cm}
\noindent{\textsf {\bf Acknowledgments:}} We acknowledge fruitful
discussions with Prof. David Chandler and Prof. Hank Ashbaugh.  SG
gratefully acknowledges partial financial support of ACS-PRF, NSF (BES
and CTS), and NSF-NSEC for Directed Assembly of Nanostructures
(DMR-0117792). TMT gratefully acknowledges financial support of the
David and Lucile Packard Foundation, the Alfred P. Sloan foundation,
and the National Science Foundation (CAREER award CTS-0448721).

\newpage
\noindent {\textbf {\textsf {REFERENCES}}} 

\newpage

\begin{figure}[h]
\caption{The lengthscale- and attraction-dependent hydrophobic 
collapse.  The PMF, $W(R_g)$, at 298 K, for the smaller C25 and larger
CG25 polymers interacting with water {\it via} attractive (LJ) and
repulsive (WCA) interactions. Conformations with $R_g<R_g^{cut}$ are
defined as compact states: $R_g^{cut}=0.6$ nm for C25, and 0.73 nm for
CG25 polymers, respectively, as shown by arrows.  PMF for C25 and CG25
polymers are zeroed at $R_g$ values of 0.83 nm and 1.4 nm,
respectively.  To make the comparison clear, the horizontal axis is
$R_g-R_g^{min}$; with $R_g^{min}= 0.385$ nm for C25 and =0.474 nm for
CG25 polymers, respectively.}
\protect\label{fig:pmfs}
\end{figure}

\begin{figure}[h]
\caption{The hydration and other contributions to PMF for polymer 
folding at 298 K.  (a) $W_{hyd}(R_g)$ contribution for C25 and CG25
polymers interacting via full LJ (filled circles), half LJ (inverted
triangles), and repulsive WCA (open circles) interactions. Panel (b)
shows the polymer-water attractive contribution, $\left< U_{pw}(R_g)
\right>$. Lines are guide to the eye. Panel (c) shows the intra
polymer $W_{vac}(R_g)$ contribution for LJ and WCA versions of C25 and
CG25 polymers.  }\protect\label{fig:hydration}
\end{figure}

\begin{figure}[h]
\caption{Free energy of polymer unfolding. The free energy 
$\Delta G_u$ obtained by integrating $W(R_g)$ curve as ${\rm
exp}(-\Delta G_u/k_BT) = \frac {\int_{R_g^{cut}}^{R_g^{max}} {\rm
exp}(-W(R_g)/k_BT) dR_g} {\int_{R_g^{min}}^{R_g^{cut}}{\rm
exp}(-W(R_g)/k_BT) dR_g}$ is shown by circles.  The curves are
obtained by fitting the data to form $\Delta G_u(T) = \Delta
H_u(T_{ref})-T\Delta S_u(T_{ref})+\Delta C_{p-u}[(T-T_{ref})-T {\rm
ln}(T/T_{ref})]$ shown by solid lines, where $T_{ref}=298$ K was
used.}\protect\label{fig:dguf}
\end{figure}

\begin{figure}[h]
\caption{Hydration contribution to the free energy and entropy of 
polymer unfolding.  (a) $\Delta G_u^{hyd} (T)$ estimated from
simulations (symbols) by subtracting appropriately integrated
polymer-water and intrapolymer energy and entropy contributions: LJ
polymers (filled circles), WCA polymers (open circles), with red and
blue indicating C25 and CG25 polymers.  Intrapolymer contributions
were obtained from vacuum runs of the polymers at the same
temperature.  Lines in panel (a) are predictions of the area model,
$\Delta G_u^{hyd}=\gamma (T) \Delta A_u$, where $\Delta A_u$ for C25
and CG25 are 0.73 and 1.66 nm$^2$, respectively, obtained by taking
differences of ensemble average area of folded and unfolded states,
and $\gamma (T) = \left[ \partial W_{hyd}/\partial A \right]_T$.  (b)
Comparison of $\Delta G_u^{hyd}$ predicted using the area-volume model
(lines) for C25 and CG25 models with simulation data (symbols). Panels
(c) and (d) show the relative conformational hydration free energy,
$\Delta W_{hyd}$, plotted as $W_{hyd}(R_g)- W_{hyd}(R_g^{cut})$, as a
function of solvent excluded volume (SEV) and SASA, respectively. SEV
and SASA are plotted relative to their average values for $R_g^{cut}$
conformers, indicated by $\Delta V$ and $\Delta A$.  Dashed lines in
panels (c) and (d) with respective slopes $c$(298 K) and $\gamma$(298
K) (values indicated) bracket the observed behavior for different
lengthscales. Panel (e) shows the hydration contribution to entropy of
unfolding. Entropy of hydration for WCA-methane obtained using test
particle insertion calculations
\protect\cite{garde:99:bpc} is also shown.}\protect\label{fig:entropy}
\end{figure}

\begin{figure}[h]
\caption{Water density fluctuations and weak dewetting of polymers. 
(a) and (b) The local water density in the vicinity of C25 and CG25
polymers, respectively, calculated as average number of hydration
waters divided by average SASA, $\left< N_w \right> / {\rm
SASA}$. These values are normalized to 1 in the wet extended states of
the polymers. (c) and (d) A measure of hydration shell compressibility
or fluctuations obtained as $\sigma^2_{N_w}/\left< N_w \right> ^2 {\rm
SASA}$, where $\sigma^2_{N_w}$ is the variance of water number
fluctuations in the hydration shell of the polymer.  A water molecule
is considered to be in the hydration shell if it is less than a
certain cut off distance (0.6 nm for C25 and 0.7 nm for CG25) from any
monomer of the polymer.}
\protect\label{fig:structure}
\end{figure}

\newpage
\begin{center}
  \includegraphics{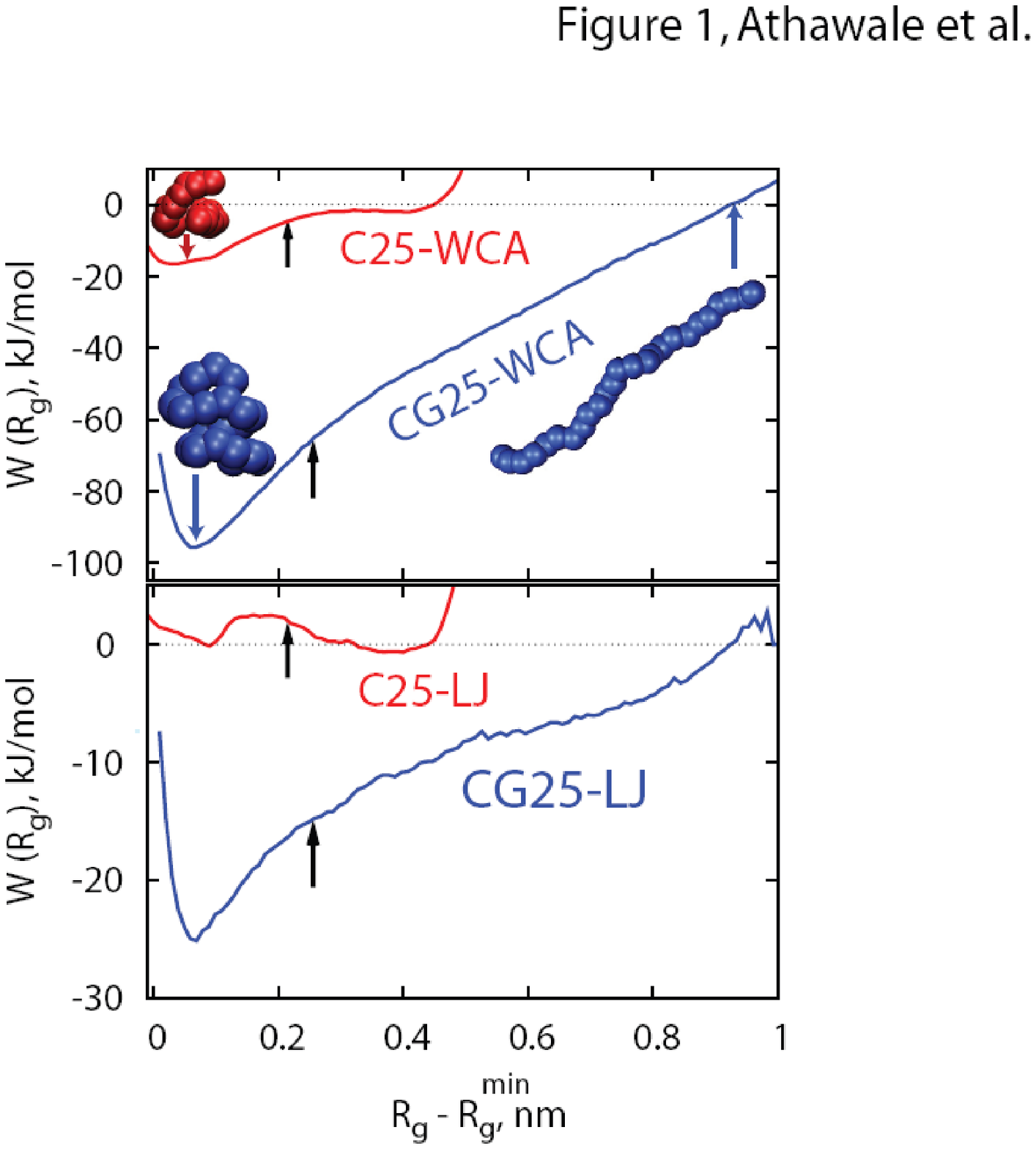}
\end{center}

\begin{center}
  \includegraphics{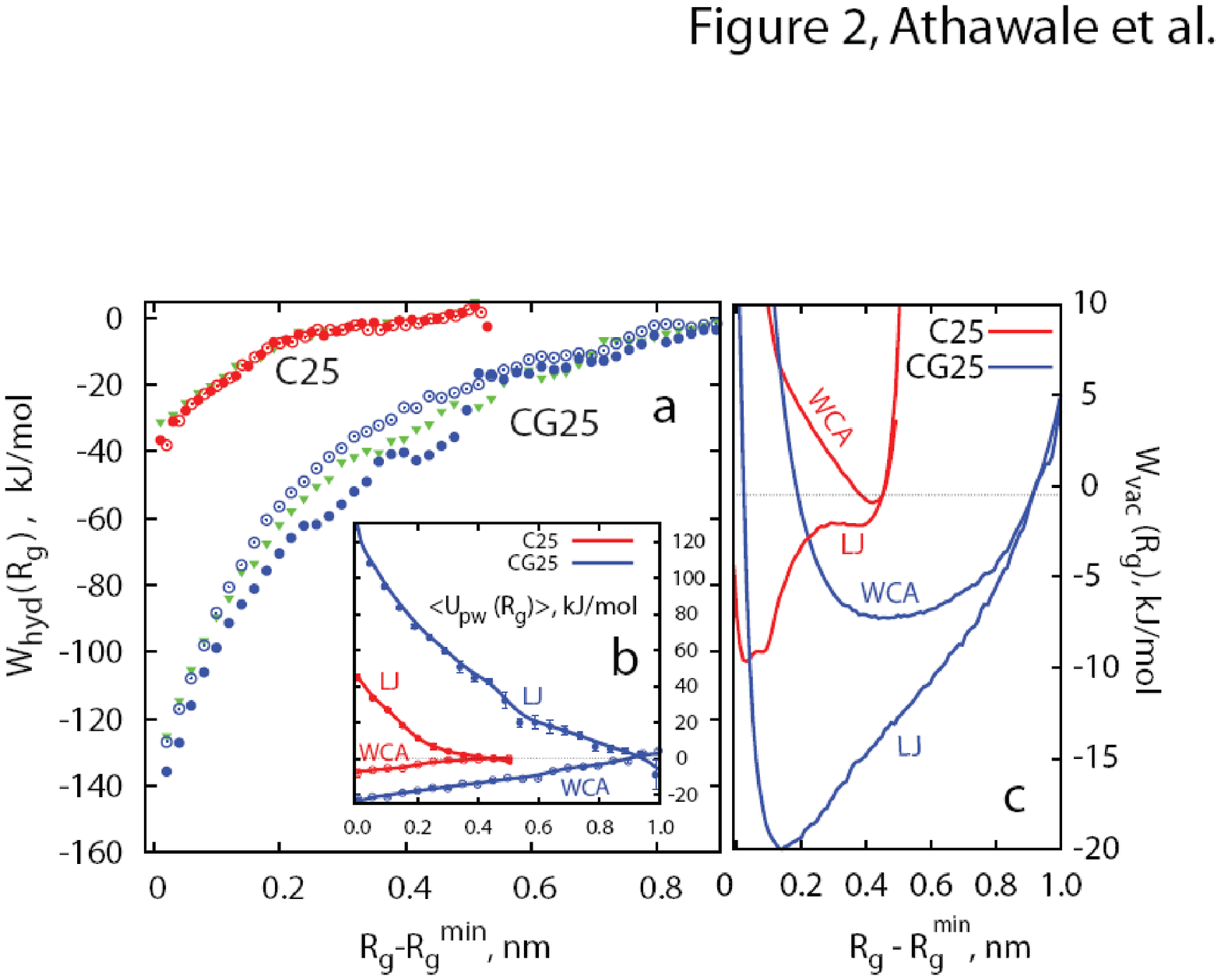}
\end{center}

\begin{center}
  \includegraphics{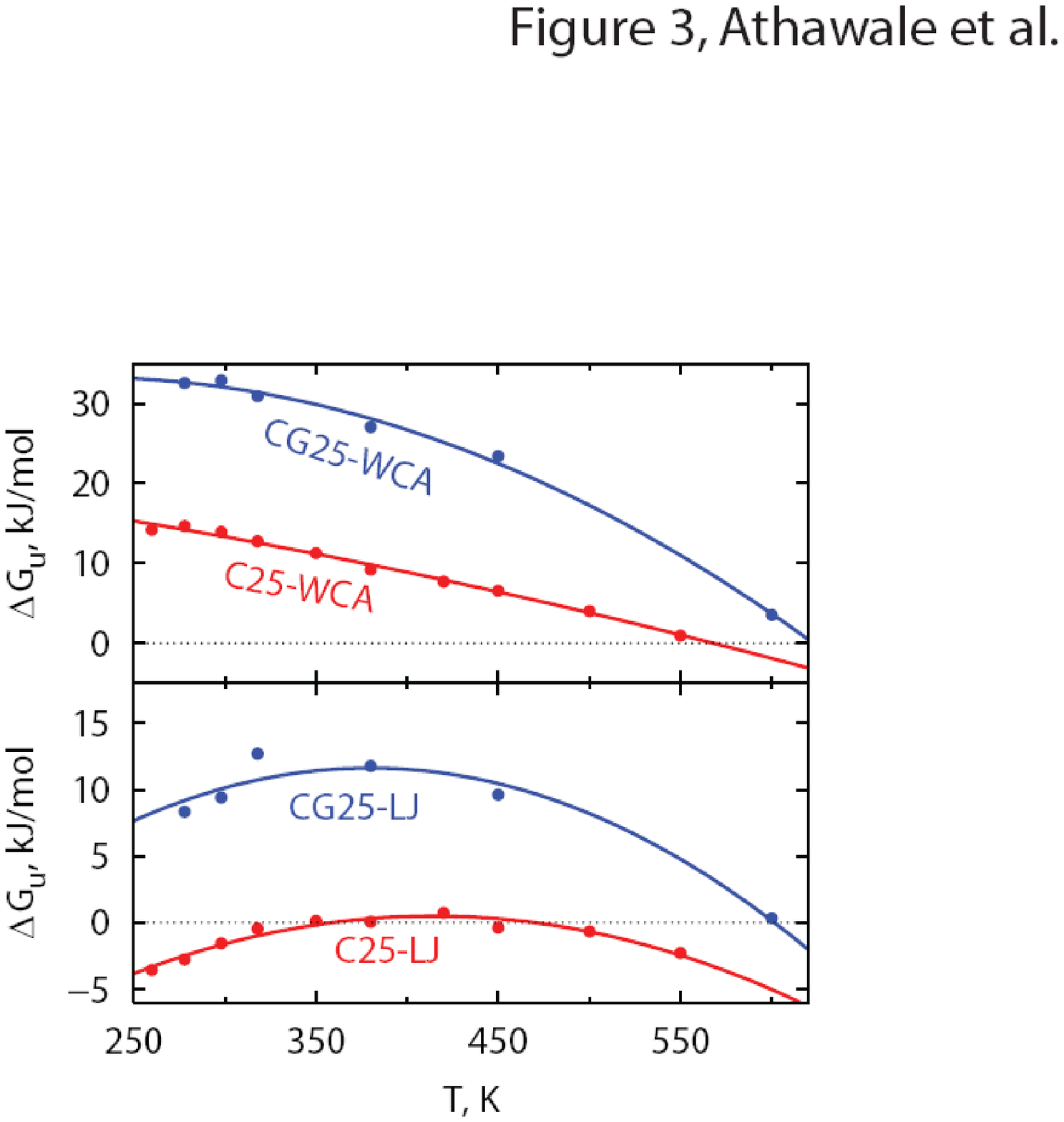}
\end{center}

  \includegraphics{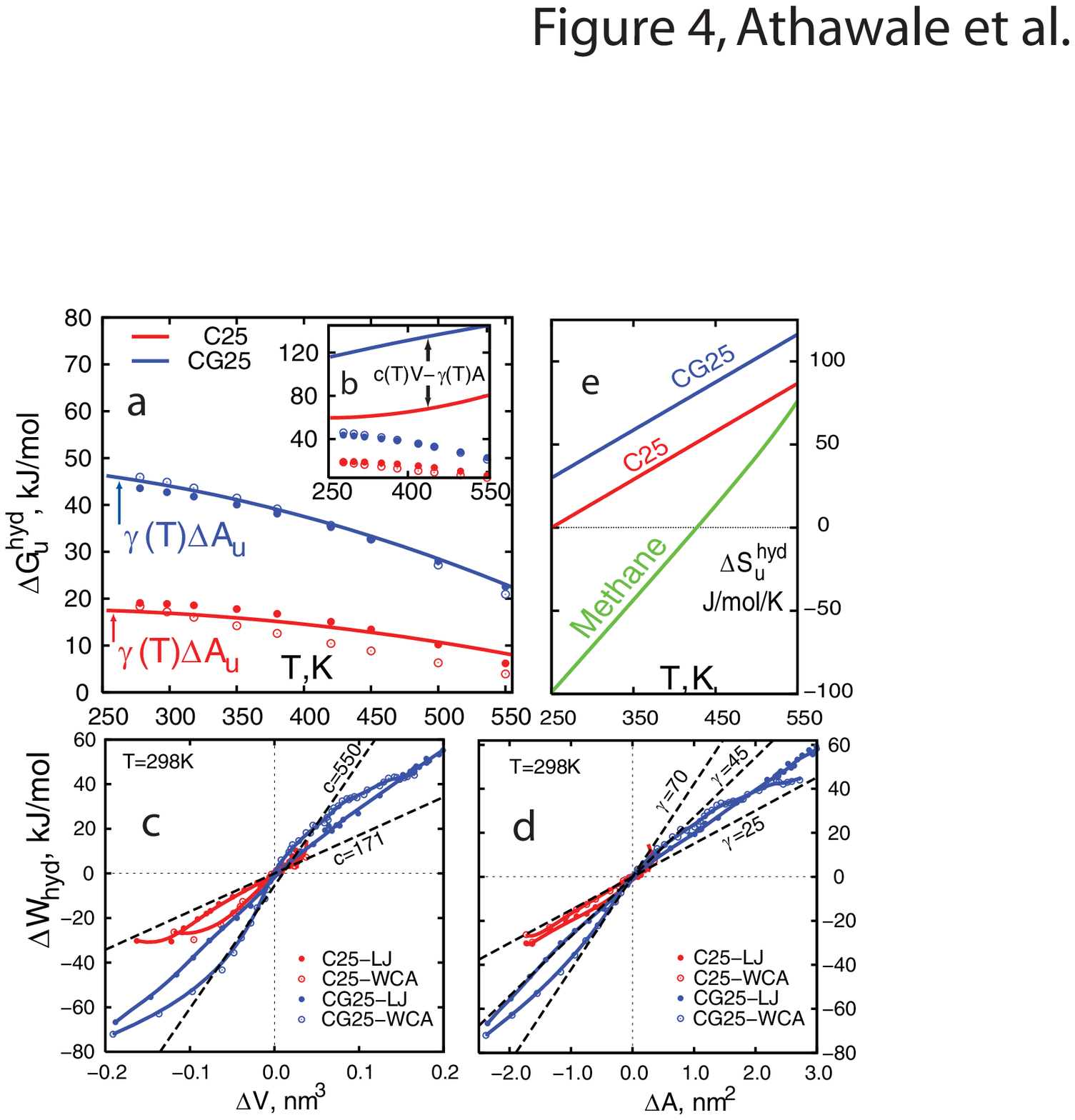}

  \includegraphics{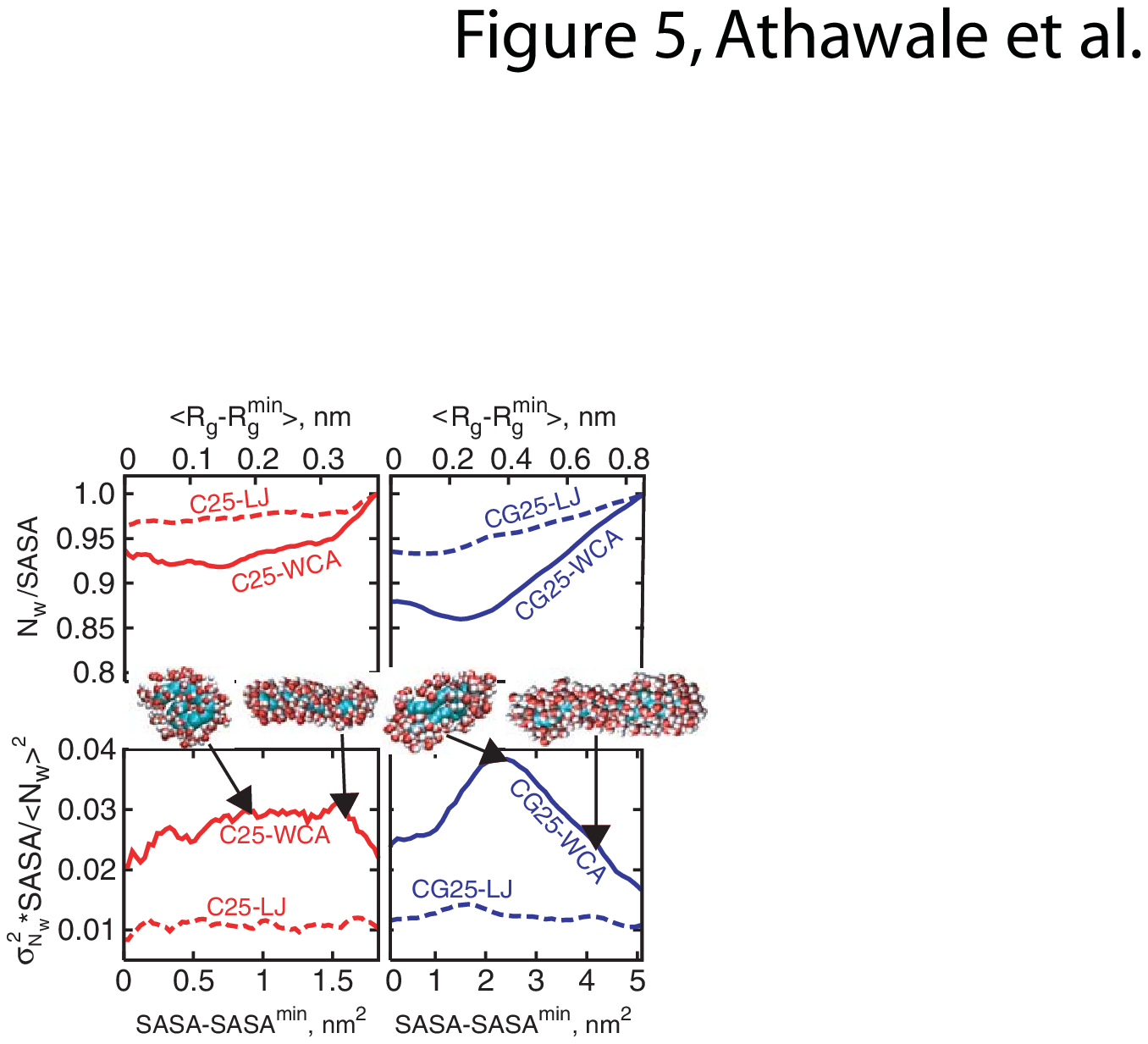}

\end{document}